\documentstyle[aps,prl,epsfig,multicol]{revtex}
\begin{document}
\title{Schwinger-Keldysh semionic approach for quantum
spin systems}

\author{M.N.Kiselev and R.Oppermann}
\address{Institut f\"ur Theoretische Physik, Universit\"at W\"urzburg,
D-97074 W\"urzburg, Germany}
\date{\today}
\twocolumn[\hsize\textwidth\columnwidth\hsize\csname@twocolumnfalse\endcsname
\maketitle

\begin{abstract}
We derive a path-integral Schwinger-Keldysh approach for quantum spin
systems.
This is achieved by means of a semionic representation of spins
as fermions with imaginary chemical potential.
The major simplifying feature in comparison with other representations
(Holstein-Primakoff, Dyson-Maleev, slave bosons/fermions etc) is
that the local constraint is taken into account exactly.
As a result, the standard diagram technique with usual Feynman codex
is constructed. We illustrate the application of this technique
for the N\'eel and spin-liquid states of the AFM Heisenberg model.\\
\mbox{}
\end{abstract}
]

Since a long time \cite{holstein} physicists are aware of the fact
that spin operators which commute on different sites and anticommute on the
same site are neither Fermi nor Bose operators. Less convergent opinions
exist on whether fermionizations or bosonizations or none of those should be
used to take care of spin statistics in many body quantum theory. At least
the answers appear to be linked to the kind of physical problem
considered. Widely accepted is the view that path integral representations
and diagrammatic expansions for spin systems are thus substantially
more complicated than those of pure fermion/boson systems.
Many variants of diagram technique \cite{book1} that are based on
different representation of spins such as Bose
\cite{holstein}-\cite{chubukov}, Fermi \cite{affleck}-\cite{vlp},
Majorana \cite{book2},
supersymmetric \cite{pepin} or Hubbard \cite{hubbard} operators
have been proposed. Another method to treat spin Hamiltonians is
based on direct representation of coherent states for spins
(nonlinear $\sigma$ model, see e.g \cite{book2}). Some of these
techniques \cite{holstein}-\cite{hubbard}, being applicable only at
low temperatures or in large spin ($S$$\gg$$1$) limit, describe
nevertheless well excitations in ordered magnets (ferromagnetic
and antiferromagnetic magnons), but fail to provide rigorous
calculations in strongly correlated systems such as Kondo lattices
or quantum magnets. Other techniques, based on a successful choice
for the hierarchy of coupling constants are mainly restricted to
equilibrium situations. The fundamental problem which is at the
heart of the difficulty is the local constraint. On one hand,
any representation of spin operators as bilinear combination of
Fermi or Bose quasiparticles makes the dimensionality of the
Hilbert space, where these operators act, greater than the
dimensionality of the Hilbert space for spin operators. As a
result, the spurious unphysical states should be excluded from the
consideration resulting in substantial complication of
corresponding rules of diagrammatical summation. On the other
hand, there is no Wick theorem directly for spin operators but the
Gaudin theorem \cite{god} instead (see also \cite{vlp,hubbard}).
It can however not avoid complications in diagram techniques based
on Hubbard operators, rendering the resummation of diagram series
in many cases practically uncontrollable. The exclusion of
double-occupied and empty states for impurity spins interacting
with a conduction electron bath (single impurity Kondo problem)
\cite{abrikos} is cured by an infinite chemical potential for
Abrikosov pseudofermions.
It works for dilute spin
subsystems, where all spins can be considered independently.
Unfortunately, attempts to generalize this technique to the
lattice of spins result in the replacement of the
local constraint (the number of particles on each site is fixed)
by a so-called global constraint (in the saddle point
approximation), where the number of particles is fixed only as an
average value for the whole crystal. There is no reason to believe
that such an approximation is a good starting point for the
description of strongly correlated systems. Besides, it is very
difficult to take into account the fluctuations related to the
replacement of a local constraint by a global one.

An alternative approach for spin Hamiltonians free of the local
constraint problem has been proposed in the pioneering paper of
Popov and Fedotov (PF) \cite{popov}. Based on  exact
fermionic representation for $S$$=$$1/2$ and $S$$=$$1$ operators, where
fermions are treated as quasiparticles with imaginary chemical
potential, these authors demonstrated the power and simplification
of the corresponding Matsubara diagram technique. For these two
special cases the Matsubara frequencies are $\omega_n$$=$$2\pi$$T$$(n+1/4)$
for $S$$=$$1/2$ and $\omega_n$$=$$2\pi$$T$$(n+1/3)$ for $S$$=$$1$
providing a rigorous description of (and restricted to) the
equilibrium situation. The semionic representation used by PF is neither
fermionic, nor bosonic, but reflects the fundamental Pauli nature of
spins. Later the generalization of the PF technique for
arbitrary spin \cite{opper} was derived by introducing proper chemical
potentials for spin fermions. The goal of this paper
is to derive a method for nonequilibrium systems, which allows to treat
quantum spin Hamiltonians on the same footing as Fermi or Bose systems.

A long time ago Keldysh \cite{keldysh} proposed a novel approach
for the description of kinetic phenomena in metals. This approach
was found especially fruitful for normal metals\cite{smith}, and, in many
recent applications, for superconductors \cite{larkin}, for
disordered interacting (normal or superconducting) electron liquids
\cite{kamenev} for example. The previous application of the real-time
formalism to the quantum theory of Bose-Einstein condensation (BEC)
\cite{stoof}
allowed the derivation of a Fokker-Planck equation, which describes both
kinetic and coherent stages of BEC. Moreover \cite{lozano} developed the
closed-time path integral formalism for aging effects in quantum
disordered systems being in contact with an environment.
The  Keldysh formalism in application to disordered systems (see \cite{kree}
-\cite{sompol}) also attracted interest some time ago as an
alternative approach to the replica technique. The main advantage
of closed-time contour calculations is an automatic normalization
(disorder independent) of the partition function. In this paper we
derive the Keldysh formalism for quantum spin systems (e.g. Heisenberg
clean and disordered magnets, Kondo lattices), which is based on
Popov-Fedotov ideas of semionic representation.

We reformulate the PF-concept adopting it to real-time formalism.
As an example we consider $S=1/2$. As it was first shown in
\cite{popov}, the partition function of a spin system with
Hamiltonian $H_S$ can be replaced by the partition function of an
effective "fermionic" system with Hamiltonian $H_F$ as follows
\begin{equation}
Z_S=Tre^{-\beta \hat H_S}=
(\pm i)^N Tr e^{-\beta(\hat H_F\pm i\pi \hat N_F/2\beta)},
\label{zf}
\end{equation}
where $\beta$$=$$1/T$ and usual "fermionic" representation of spin similar to
e.g. Abrikosov pseudofermions \cite{abrikos} is used:
$S^+$$=$$f^\dagger_\uparrow f_\downarrow,$
$S^-$$=$$f^\dagger_\downarrow f_\uparrow,$
$S^z$$=$$\frac{1}{2}$$(f^\dagger_\uparrow
f_\uparrow$$-$$f^\dagger_\downarrow f_\downarrow),$
and $N_F$$=$$f^\dagger_\uparrow f_\uparrow$$+$$f^\dagger_\downarrow f_\downarrow$.

Representing spins as bilinear combinations of Fermi operators,
we enlarged by a factor of two the Hilbert space of the Hamiltonian.
In addition to physical states $|1,0\rangle$ and $|0,1\rangle$
two unphysical states $|1,1\rangle$ and $|0,0\rangle$ are introduced.
Nevertheless in the average over all states unphysical states cancel each
other, since
$Tr_{unphys}(\exp(\mp i\pi/2))^{N_{F}}=(\mp i)^0+(\mp i)^2=0.$
This representation being of semionic origin results in conventional Matsubara
diagram technique with
$\omega_n=2\pi T(n+1/4)$ or $\omega_n=2\pi T(n+3/4)$ \cite{bose}
depending on the sign in expressions (\ref{zf}).
Besides, one can introduce the auxiliary distribution function for
quasiparticles \cite{op2}
\begin{equation}
f^{(1/2)}(\epsilon)= T\sum_n\frac{e^{i\omega_n\tau|_{+0}}}{i\omega_n-\epsilon}=
\frac{1}{e^{\pm i\pi/2}\exp(\beta\epsilon)+ 1}
\label{d12}
\end{equation}
where signs $\pm$ in the exponent (\ref{d12}) are the same as in (\ref{zf}).
We note that,
since auxiliary Fermi fields do not represent the true quasiparticles of the
problem helping only to treat properly the spin operators,
the distribution function for these objects in general should not be a
real function, e.g.
$f^{(1/2)}=n(2\epsilon)\mp \frac{i}{2}{\rm sech}(\epsilon/T)$
where $n(x)=[\exp(x/T)+1]^{-1}$ is the standard Fermi distribution function.
As we shall see for arbitrary value of spin,
$1- 2 {\it Re}f^{(S)}(\epsilon)=B_S(\epsilon/T)$
is expressed in terms of the Brillouin function
$B_S(x)=(1+\frac{1}{2S})\coth((1+\frac{1}{2S})x)
-\frac{1}{2S}\coth(\frac{x}{2S})$, e.g. for $S=1/2$, $B_{1/2}(x)=\tanh(x)$.
We also note that in the $T\to 0$-limit the imaginary part of $f^{(1/2)}$
satisfies the identity ${\it Im}f^{(1/2)}(x)=\mp i\pi T\delta(x)/2$.

The spin correlation functions of any order can be expressed in terms of
the two-component field $\psi^T$$=$$(f_\uparrow\;f_\downarrow)$:
$$
\langle S^{\alpha_1}_{i_1}(t_1)...S^{\alpha_n}_{i_n}(t_n)\rangle
=Tr(\rho_0 (\psi^\dagger\sigma^{\alpha_1}_{i_1}\psi)_{t_1}...
(\psi^\dagger\sigma^{\alpha_n}_{i_n}\psi)_{t_n})
$$
where $\rho_0=\exp(-\beta H_0)/Tr\exp(-\beta H_0)$
is the density matrix and $\sigma$ denotes Pauli matrices.
We included the term $i\pi N_{F}/(2\beta)$ into the Hamiltonian
$H_0=-h\sum_iS^z_i\pm i\pi T/2\sum_i N^{(i)}_{F}$ of
noninteracting spins in a uniform external magnetic field $h$,
since it exists both in numerator and denominator of $\rho_0$.
\vspace*{-3mm}
\begin{figure}
\begin{center}
\epsfig{%
file=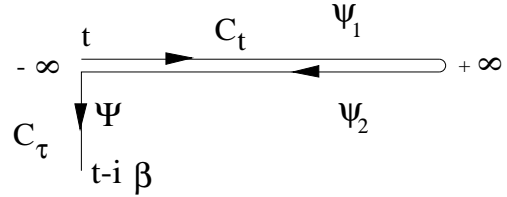,%
figure=figure1.eps,%
height=25mm,%
width=65mm,%
angle=0,%
}
\end{center}
\caption{Keldysh double side  contour going along  real time axis
$-\infty \to +\infty \to -\infty$ and "closed" in imaginary time.}
\label{cont}
\end{figure}
Following the standard route \cite{babich} we can express the partition
function of the problem containing spin operators as a path integral over
Grassmann variables $\bar \psi$,$\psi$
\begin{equation}
{\cal Z}/{\cal Z}_0=\int D\bar\psi D\psi\exp(i{\cal A})/
\int D\bar\psi D\psi\exp(i{\cal A}_0)
\label{pfunk}
\end{equation}
where actions ${\cal A}$ and ${\cal A}_0$ are taken as an integral along the
closed-time contour $C_t+C_\tau$ which is shown in Fig.1.
The contour is closed at $t=-\infty+i\tau$ since
$\exp(-\beta H_0)=T_\tau\exp\left(-\int_0^\beta H_0 d\tau\right).$
We denote the $\psi$ fields on upper and lower sides of the contour
$C_t$ as $\psi_1$ and $\psi_2$ respectively. The fields $\Psi$ stand for
the contour $C_\tau$. These fields provide matching conditions for
$\psi_{1,2}$ and are excluded from final expressions.
Taking into account the semionic boundary conditions for generalized Grassmann
fields
$\Psi^\mu(\beta)$$=$$i\Psi^\mu(0)$,
$\bar\Psi^\mu(\beta)=-i\bar\Psi^\mu(0)$
one gets the matching conditions for $\psi_{1,2}$ at $t=\pm\infty$.
\begin{equation}
\psi^\mu_1(-\infty)=i\psi^\mu_2(-\infty)e^{\beta h\sigma^z_\mu},\;\;\;
\psi^\mu_1(+\infty)=\psi^\mu_2(+\infty)
\label{bc2}
\end{equation}
The correlation functions can be represented as functional derivation
of the generating functional
$$
Z[\eta]/{\cal Z}_0=\int D\bar\psi D\psi\exp\left(i{\cal A}+
i\oint_C d t (\bar\eta\sigma^z\psi+\bar\psi\sigma^z \eta)\right)
$$
where $\eta$ represents sources and $\sigma^z$ matrix stands for "causal"
and "anti-causal" orderings along the contour.

The on-site Green's functions (GF) which are matrices $4\times 4$ with
respect to both Keldysh (lower) and spin (upper) indices are given by
$$
G_{\mu\nu}^{\alpha\beta}(t,t')=
-i\frac{\delta}{i\delta\bar \eta_\mu^\alpha(t)}
\frac{\delta}{i\delta \eta_\nu^\beta(t')}
Z[\eta]|_{\bar\eta,\eta\to 0}
$$
After a standard shift-transformation \cite{babich} of fields $\psi$
the Keldysh GF of free PF-fermions assumes the form
\begin{eqnarray}
G_0^\alpha(\epsilon)=G^{R,\alpha}_0
\left(
\begin{array}{cc}
1 - f_\epsilon &  -f_\epsilon\\
1 - f_\epsilon &  -f_\epsilon
\end{array}\right)-
G^{A,\alpha}_0
\left(
\begin{array}{cc}
-f_\epsilon & -f_\epsilon\\
1 - f_\epsilon & 1 - f_\epsilon
\end{array}
\right)
\nonumber
\end{eqnarray}
where the retarded and advanced GF's are
\begin{equation}
G^{(R,A)\alpha}_0(\epsilon)=(\epsilon +\sigma^z_\alpha h/2 \pm i\delta)^{-1},\;\;\;\;
f_\epsilon=f^{(1/2)}(\epsilon).
\label{gra}
\end{equation}
The interdependence of matrix elements of the GF in Keldysh-space is more
transparent after rotation
\begin{eqnarray}
\hat G\Rightarrow \frac{1-i\sigma^y}{\sqrt{2}}\sigma^z G
\frac{1+i\sigma^y}{\sqrt{2}}
=\left(
\begin{array}{cc}
G^R & G^K\\
0 & G^A
\end{array}
\right)
\label{triang}
\end{eqnarray}
where $G_0^K$$=-i2\pi\delta(\epsilon\pm h/2)\left[B_{1/2}(\epsilon/T)\pm
i\hspace{.1cm}{\rm sech}(\epsilon/T)\right]$.
We emphasize, that unlike diagrammatic techniques for Fermi and Bose operators,
the off-diagonal element (Keldysh-component) in semionic representation is
expressed in terms of Brillouin function, containing correct information
about occupied states.
We recall that diagonal elements of the matrix (\ref{triang}) in "triangular"
representation satisfy the Dyson equation providing the exact description
of the system. The equation of motion for $G^K$ generally constitutes
the quantum-kinetic equation.

Let us illustrate the application of the Schwinger-Keldysh formalism for spin
Hamiltonians. We consider the Heisenberg model with nearest neighbour
interaction
$$
H_{int}=-\sum_{<ij>}J_{ij}\left(\vec{S}_i\vec{S}_j-\frac{1}{4}\right)=
\frac{1}{2}\sum_{<ij>}J_{ij}\psi^\dagger_i\psi_j\psi^\dagger_j\psi_i
$$
We firstly discuss the N\'eel solution for the Heisenberg
model with isotropic antiferromagnetic (AFM)
exchange ($J<0$). Applying the
PF-transformation to the partition function one obtains the action as an
integral along the closed-time Keldysh-contour
\begin{equation}
{\cal A}={\cal A}_0+{\cal A}_{int}={\cal A}_0+
\oint_{C}d t\sum_{\bf q}J({\bf q})\vec{S}_{\bf q}(t)\vec{S}_{-\bf q}(t)
\label{abeg}
\end{equation}
where ${\cal A}_0$ corresponds to noninteracting fermions
\begin{eqnarray}
{\cal A}_0=\oint_C dt\sum_i\bar\psi_i
\left(
\begin{array}{cc}
(G^{R,\alpha}_0)^{-1} &0\\
0 & (G^{A,\alpha}_0)^{-1}
\end{array}
\right)\psi_i
\label{a0}
\end{eqnarray}
We denote $J_{\bf q}$$=$$J$$\sum_{<{\bf l}>}$$e^{i{\bf ql}}$,
$\nu_{\bf q}$$=$$J_{\bf q}$$/$$J_0$
and apply  8-component PF  representation  with
$\psi^T$$=$$(\tilde\psi^T_{\bf k}$$\tilde\psi^T_{\bf k+Q})$,
where ${\bf Q}=(\pi,...,\pi)$ for hypercubic lattice.
To decouple the four-fermion term along the Keldysh-contour
with the help of Hubbard-Stratonovich transformation we introduce the
two-Keldysh-component {\it vector} (Bose) field
$\vec{\Phi}^T=(\vec{\Phi}_1\;\vec{\Phi}_2)$.
As a result we obtain
\begin{equation}
{\cal A}_{int}=-\frac{1}{2}Tr(\vec{\Phi}_{\bf q}^T
J^{-1}_{\bf q}\sigma^z \vec{\Phi}_{\bf q})
+Tr(\bar\psi\vec{\Phi}\vec{\sigma}\gamma^\mu\psi)
\label{aint}
\end{equation}
Now we integrate out $\psi$ fields and express the effective
action in terms of $\vec\Phi$ fields
$$ {\cal A}_{eff}=
-\frac{1}{2}Tr(\vec{\Phi}_{\bf q}^T J^{-1}_{\bf q}\sigma^z \vec{\Phi}_{\bf q})
+Tr\ln\left(G_0^{-1}+\vec{\Phi}_\mu\vec{\sigma}\gamma^\mu\right)
$$
where $\gamma^\mu$$=$$(\sigma^z\pm 1)/2$ acts in Keldysh space.
Since in general $\vec\Phi$ is a time- and space-dependent
fluctuating field the partition function (\ref{pfunk}) cannot be evaluated
exactly. Nevertheless, when a magnetic instability occurs, we can
represent the longitudinal component of this field as a
superposition of a staggered time-independent part
("staggered condensate") and a fluctuating field
\begin{equation}
\Phi^z_\mu({\bf q},\omega)={\cal N}J_{\bf q}\gamma^\mu\delta_{\bf
q,Q} \delta(\omega)+\phi^z_\mu({\bf q},\omega),
\label{cond1}
\end{equation}
where ${\cal N}$ is a staggered magnetization
and $\Phi_\mu^{\pm}({\bf q},\omega)=\phi_\mu^\pm({\bf q},\omega)$ with
the matching conditions at $t=\pm\infty$
involving the dispersions of excitations $\omega_{\bf p}$
\begin{equation}
\phi_1^{\pm}(-\infty)=\phi_2^{\pm}(-\infty)e^{i\beta\omega_{\bf p}},\;\;
\phi_1^{\pm}(+\infty)=\phi_2^{\pm}(+\infty).
\label{bc3}
\end{equation}
We expand
$Tr\ln(G_0^{-1}+\vec{\phi}_\mu\vec{\sigma}\gamma^\mu)$ in accordance with
\begin{equation}
Tr\ln(...)=Tr\ln G_0^{-1}+\sum_{n=1}^{\infty}
\frac{(-1)^{n+1}}{n}(G_0\vec{\phi}_\mu\vec{\sigma}\gamma^\mu)^n
\label{tr}
\end{equation}
\vspace*{-5mm}
\begin{figure}
\begin{center}
\epsfxsize6cm
\epsfbox{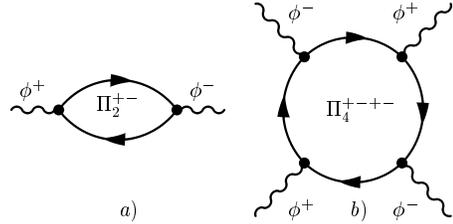}
\caption{Feynman diagrams contributing to dispersion (a) and  damping (b) of
magnons. Solid line denotes PF fermions.}
\end{center}
\end{figure}
\vspace*{-5mm}
The spectrum of the excitations (AFM magnons) can be defined as poles
of the transverse GF
$D^{+-}_{{\bf x},t}$$=$$D({\bf x},t)$$=$$-i
\langle T_C\phi_1^+({\bf x},t)\phi_1^-(0,0)\rangle$.
The procedure of the calculation of this GF is similar to that for a
"fermionic" GF.
Introducing the sources and evaluating (\ref{tr}) one gets
\begin{eqnarray}
D_0(\omega)=D^{R}_0
\left(
\begin{array}{cc}
1+N_\omega & N_\omega\\
1+N_\omega & N_\omega
\end{array}\right)-
D^{A}_0
\left(
\begin{array}{cc}
N_\omega & N_\omega\\
1+N_\omega  & 1+N_\omega
\end{array}
\right)
\nonumber
\end{eqnarray}
where the retarded and advanced magnons GF's are
$$
D^{R,A}({\bf p},\omega)=(\omega-\omega({\bf p})\pm i\delta)^{-1},\;\;
N_\omega=(\exp(\beta\omega)-1)^{-1}
$$
The magnon spectrum is determined by the zeros of the expression
$J^{-1}_{\bf q}-\Pi_2^{+-}(\omega)$ (see Fig.2a) in equilibrium
\begin{equation}
\omega_{\bf p}=|J_0|{\cal N}\sqrt{1-\nu^2_{\bf p}}\Rightarrow c|{\bf p}|,\;
{\cal N}=\tanh\left(\frac{J_{\bf Q}{\cal N}}{2T}\right)
\label{afm}
\end{equation}
The magnon-damping is defined by four-magnon processes
$\Pi_4^{+-+-}$, shown in Fig2.b.
The derivation of the kinetic equation and calculation of magnon damping
is reserved here for a detailed publication.
The results (\ref{afm}) (and similar for quantum FM)
are in full agreement with the spin-wave
theory (see, for example, \cite{book1} and \cite{arovas,vlp}).

The second possibility to decouple the four-fermion term in the Heisenberg
model is provided
by the bi-local {\it scalar} bosonic field $\Lambda_{ij}$
depending on two sites.
Introducing new coordinates $\vec{R}$$=$$(\vec{R}_i$$+$$\vec{R}_j)$$/2$, 
$\vec{\rho}$$=$$\vec{R}_i$$-$$\vec{R}_j$
and applying a Fourier transformation we obtain the effective action
$$ {\cal A}_{eff}=
-\frac{1}{2}Tr(\Lambda_{Pq_1}^T J^{-1}_{\bf q_1-q_2}\sigma^z \Lambda_{Pq_2})
+Tr\ln\left(G_0^{-1}-\Lambda_\mu\gamma^\mu\right)
$$
This effective action describes the nonequilibrium quantum spin-liquid (SL).
We confine ourselves to consider the uniform phase \cite{ioffe} of
Resonant Valence Bonds.
It is sutable to rewrite the functional in new variables, namely
amplitude $\Delta$  and phase $\Theta$$=$$\vec{\rho}$$\vec{A}(\vec{R})$,
according to formula
\begin{equation}
\Lambda^{<ij>}_\mu(\vec{R},\;\vec{\rho})=
\Delta(\vec{\rho}) J\gamma^\mu\exp\left(i\vec{\rho}\vec{A}_\mu(\vec{R})\right)
\label{uni}
\end{equation}
The exponent in (\ref{uni}) stands for gauge fluctuations to be taken in
eikonal approximation.
The spectrum of excitations in the uniform SL
is defined by the zeros of
${\Large \pi}^{R,\alpha\beta}_{q,\omega}$$=$$Tr(p^\alpha p^\beta
(G^R_{p+q}G^K_p$$+$$G^K_{p+q}G^A_p)$$+$$\delta_{\alpha\beta}
f(J_{\bf p}\Delta))$ in equilibrium \cite{reizer}
and is purely diffusive (see e.g.\cite{ioffe})
\begin{equation}
\omega=iJ\Delta |{\bf q}|^3,\;\;\;\;
\Delta=-\sum_{\bf q}\nu({\bf q})\tanh\left(\frac{J_{\bf q}\Delta}{T}\right)
\label{rvb}
\end{equation}
The quantum kinetic equation for nonequilibrium spin liquids can be
obtained by taking into account the higher order diagrams
similarly to Fig.2b with current-like vertices and will be presented elsewhere.

We discuss finally the Schwinger-Keldysh formalism for spins $S>1/2$.
As shown by Popov and Fedotov for $S$$=$$1$, it is possible to eliminate the
unphysical states by introducing three-component fermions
$\psi^T$$=$$(f_\uparrow\;f_0\;f_\downarrow)$ with imaginary chemical potential
$\lambda$$=$$-i\pi$$T/3$. The boundary conditions for $\Psi$ on the imaginary
part of the contour $C_\tau$ read as follows
$\Psi^\mu(\beta)$$=$$e^{i\pi/3}\Psi^\mu(0)$,
$\bar\Psi^\mu(\beta)$$=$$e^{-i\pi/3}\bar\Psi^\mu(0).$
As a result, the distribution function in equilibrium is
$f^{(1)}(\epsilon)$$=$$1$$/$$\left[e^{\pm i\pi/3}\exp(\epsilon/T)+1\right].$
Thus the Schwinger-Keldysh formalism with $6\times 6$ matrices for GF
(\ref{triang}) and $f_\epsilon$$=$$f^{(1)}(\epsilon)$ in equilibrium is
obtained. The off-diagonal  Keldysh component is given by
$$G_0^K=-i2\pi\delta(\epsilon\pm h)
[B_1(\epsilon/T) \pm i\sqrt{3}
\sinh(\frac{\epsilon}{2T})/\sinh(\frac{3\epsilon}{2T})].$$
For arbitrary spin values $S$$>$$1$ there is no unique imaginary chemical
potential for $2$$S$$+$$1$ component PF-fermions, but
instead they are distributed on each lattice site $j$ according to
$$
P(\lambda_j)=\sum_{l=0}^{[S-1/2]}a_l\delta(\lambda_j-\lambda_l),\;\;
a_l=\frac{2i}{2S+1}\sin\left(\pi\frac{2l+1}{2S+1}\right),
$$
where $\lambda_l$$=$$i\pi$$T(2l+1)$$/$$(2S+1)$ \cite{opper}.
Thus, the Schwin\-ger-Kel\-dysh approach can
be generalized for arbitrary spin values in the same fashion as for
$S$$=$$1/2$ and $S$$=$$1$.

Summarizing, we derived the technique applicable for nonequilibrium
dynamics of quantum spin systems.
Unlike other techniques this approach takes into account the constraint
rigorously and allows one to treat spins on the same footing as Fermi and
Bose systems. The method derived can be applied especially to
problems where the local constraint becomes important, e.g.
quantum phase transition in clean and disordered magnets, spin glasses,
Kondo lattices, nonequilibrium Kondo systems etc.

We acknowledge useful discussions with  L.V.Keldysh and many helpful
conversations with H.Feldmann and K.Kikoin.
This work is supported by the SFB410 (II-VI semiconductors).
MNK express his gratitude to Alexander von Humboldt Foundation for
support of his research.


\begin{thebibliography} {99}
\bibitem{bose} Formally one can also derive real-time semionic approach
for spin operators starting from hard-core bosons with complex chemical potential.
Nevertheless, this representation is not significantly different
from Abrikosov's approach  [6] and therefore
is not convenient for calculations.
\end{thebibliography}
\end{document}